\documentclass{PoS}
\usepackage{graphicx}

\title{Scattering states in Bethe-Salpeter equation}

\ShortTitle{Scattering states in Bethe-Salpeter equation}

\author{\speaker{V.A.~Karmanov}\\
        Lebedev Physical Institute,\\
        Leninsky prospect 53, 119991 Moscow, Russia\\
        E-mail: \email{karmanov@sci.lebedev.ru}}
\author{J.~Carbonell\\
Institut de Physique Nucl\'eaire, \\
Universit\'e Paris-Sud, IN2P3-CNRS, 91406 Orsay Cedex, France\\
        E-mail: \email{carbonell@ipno.in2p3.fr}}

\abstract{The off-mass shell scattering amplitude, satisfying the Bethe-Salpeter equation for spinless particles in Minkowski space with the ladder kernel, is computed for the first time.}

\FullConference{XXI International Baldin Seminar on High Energy Physics Problems \\
                 September 10-15, 2012\\
                 JINR, Dubna, Russia}

\begin{document}

\section{Introduction}
\label{intro}
The inhomogeneous Bethe-Salpeter (BS) equation in Minkowski space \cite{bs}
provides a covariant four-dimensional  description of two-body  scattering states.
In the case of scalar particles it has the form:
\begin{equation}\label{BSE}
F(p,p_s;P)=K(p,p_s;P)- i\int\frac{d^4p'}{(2\pi)^4}
\frac{K(p,p';P) F(p',p_s;P)}
{\left[\left(\frac{1}{2}P+p'\right)^2-m^2+i\epsilon\right]
\left[\left(\frac{1}{2}P-p'\right)^2-m^2+i\epsilon\right]}
\end{equation}

The kernel $K$  for the one-boson exchange model is given by:
\begin{equation}\label{obe}
K(p,p';P)=-\frac{g^2}{(p-p')^2-\mu^2+i\epsilon}
\end{equation}
We introduce the coupling constant $\alpha$ related to $g$ by:
$$
g^2=16\pi m^2\alpha
$$
and use the partial wave decomposition according to \cite{IZ}:
$$
F(\theta)=16\pi\sum_{l=0}^{\infty} (2l+1)F_l P_l(\cos\theta)
$$
In the center of mass frame, $\vec{P}=0$, $P_0=\sqrt{s}=2\varepsilon_{p_s} =2\sqrt{m^2+p_s^2}$ and for a given incident  momentum $p_s$,
the partial wave off-mass shell amplitude $F_l$ depends on two scalar variables $p_0$ and $|\vec{p}|$. It will be hereafter denoted by
$F_l(p_0,p;p_s)$   setting $p=|\vec{p}|$, $p_s=|\vec{p_s}|$.
We consider for simplicity the S-wave equation which reads:
\begin{eqnarray}\label{eq1a}
F_0(p_0,p;p_s)=K_0(p_0,p;p_s)&-& \frac{4i}{\pi^2}\int_0^{\infty} {p'}^2dp'\int_{-\infty}^{\infty}dp'_0
\\
&\times&\frac{K_0(p_0,p;p'_0,p') F_0(p'_0,p';p_s)}
{\Bigl({p'}_0^2+2 p'_0\varepsilon_{p_s} +{p_s}^2-{p'}^2+i\epsilon\Bigr)
\Bigl({p'}_0^2-2p_0\varepsilon_{p_s} +{p_s}^2-{p'}^2+i\epsilon\Bigr)}
\nonumber
\end{eqnarray}
with
\begin{eqnarray}\label{K0}
K_0(p_0,p;p'_0,p') &=&-\frac{1}{32\pi}\int_{-1}^{1} dz \frac{g^2}{(p_0-p'_0)^2 -(p^2-2pp'z+{p'}^2)-\mu^2+i\epsilon}
\nonumber\\
&=&-\frac{\alpha m^2}{4pp'}\log\frac{|\eta+1|}{|\eta-1|}+ \frac{i\alpha\pi
m^2}{4 pp'}U(\eta),
\end{eqnarray}
$$
U(\eta)=\left\{\begin{array}{ll} 1,  &\mbox{if $|\eta|\leq 1$}\\
0, &\mbox{if $|\eta|> 1$}
\end{array}\right.
$$
and
$$
\eta=\frac{(p_0-p'_0)^2-p^2-{p'}^2-\mu^2}{2pp'}
$$

The on-shell amplitude $F^{on}_l=F_l(p_0=0,p=p_s;p_s)$
determines the phase shift according to:
\begin{equation}\label{delta}
\delta_l=\frac{1}{2i}\log\Bigl(1+\frac{2ip_s }{\varepsilon_{p_s}} F^{on}_l\Bigr)
\end{equation}
The knowledge of this function in the entire domain of its arguments -- i.e. the off-shell amplitude -- is mandatory for some interesting physical applications, like for instance  computing  the transition e.m. form factor $\gamma^*d\to np$ or  solving the BS-Faddeev equations.
As far as we aware, this quantity  has not  been obtained until now.

The numerical solution of the BS equation in Minkowski space is complicated by the existence of singularities in the amplitude as well as in the integrand of (\ref{BSE}). These singularities are integrable in the mathematical sense, due to $i\epsilon$ in the denominators of propagators,
but  their  integration is a quite delicate task and requires the use  of appropriate analytical as well as  numerical methods.

To avoid these singularities, the BS equation was first solved in Euclidean space. These solutions provided   on-shell quantities like  binding energies and  phase shifts \cite{tjon}. However we have shown \cite{ck-trento} that the Euclidean BS amplitude cannot be used to calculate electromagnetic form factors, since the corresponding integral does not allow the Wick rotation.
One therefore needs the BS amplitude in Minkowski space.

This amplitude has been computed for a separable kernel  (see \cite{burov} and references therein).
For a  kernel given  by field theory rules --  ladder and cross ladder --   the Minkowski BS amplitude was first obtained in our preceding works \cite{bs1,bs2} in the  bound state problem.
To this aim, we developed an original method based on the Nakanishi integral representation of the amplitude.
A similar method for the scattering states has been proposed in \cite{fsv-2012} although  the numerical solutions are not yet available.

We present in this paper some details of a new method \cite{Fukuoka} providing a direct solution of the original BS equation. It is based  on a carefull
treatment of the singularities and allows to compute the corresponding off-shell scattering amplitude in Minkowski space.
We will give the low energy parameters in the case of spinless particles and ladder kernel.

\section{Method}
\label{sec1}
There are four sources of singularities in the  r.h.-side of the BS equation (\ref{BSE}) which are detailed below.

{\it (i)} The constituent propagators in (\ref{eq1a}) vs. $p'_0$ have two poles, each of them is represented  as:
$$
\frac{1}{p'_0-a-i\epsilon}=PV\frac{1}{p'_0-a}+i\pi \delta(p'_0-a)
$$
where $PV$ means the principal value. In the product of four pole terms, the only non vanishing contributions come from the product of four $PV$'s without delta-functions, from the terms with three $PV$'s and one delta-function and from the term with two $PV$'s and two delta's. After partial wave decomposition the 4D integral BS equation (\ref{BSE}) is reduced to a 2D one, eq. (\ref{eq1a}). Integrating in (\ref{eq1a}) over $p'_0$, we obtain  in addition to the 2D  part, a 1D integral over $p'$ and a non integrated term.
The singularities due to the $PV$'s are eliminated by subtractions according to the identity:
$$
PV\int_0^{\infty}\frac{f(p'_0)dp'_0}{{p'_0}^2-a^2}=
\int_0^{\infty}\left(\frac{f(p'_0)}{{p'_0}^2-a^2} - \frac{f(a)}{{p'_0}^2-a^2}\right)dp'_0
$$
The integrand in r.h.-side is not singular.

{\it (ii)} The  propagator of the exchanged particle has the pole singularities which, after  partial wave decomposition, turn into logarithmic ones, eq. (\ref{K0}). Their positions are found analytically and the numerical integration over $p'_0$ variable is split into intervals between two consecutive singularities, namely:
$$\int_0^{\infty}[ \ldots ] \;  dp'_0= \int_0^{sing_1} [ \ldots ] \; dp'_0 +
\int_{sing_1}^{sing_2}[ \ldots ] \; dp'_0
+\dots
$$
Each of these integrals is made regular with an appropriate change of variable. We proceed in a similar way for the $p'$ integration.

{\it (iii)}  The inhomogeneous (Born) term is given by the ladder kernel and is also singular in both variables. The positions of these singularities are  analytically known.

{\it (iv)}  The amplitude $F_0$ itself has many singularities, among which
the strongest ones are resulted from the Born term $K_0(p_0,p;p_s)$. This makes difficult its representation on a basis of regular functions as well as its numerical integration. To circumvent this difficulties we made the replacement
$F_0(p_0,p;p_s)=K_0(p_0,p;p_s)f_0(p_0,p;p_s)$, where $f_0$ is a smooth function. After that, the singularities of the inhomogeneous term are canceled.
We obtain in this way a non-singular equation for  $f_0$ which we solve by standard methods. Then we restore the BS off-mass shell amplitude $F_0$ in Minkowski space.

\section{Numerical results}
\label{sec2}
We first applied this method to solve the bound state problem by dropping the inhomogeneous term in (\ref{BSE}). The binding energies coincide, within four-digit accuracy, with the ones calculated in our previous work \cite{bs1} and with the Euclidean space results.

Solving eq. (\ref{eq1a}), the S-wave off-shell scattering amplitude $F_0$ is calculated and the phase shifts are extracted by means of eq. (\ref{delta}).
Above the first inelastic threshold , $p^*_s(\mu)=\sqrt{m\mu+\mu^2/4}$, this phase shifts  have an imaginary part, which is also found.
By performing a Wick rotation in (\ref{eq1a}) and taking into account the contributions of singularities -- which, in contrast to the bound state case, are crossed by the rotation  contour --  we derived an Euclidean space equation similar to one obtained in \cite{tjon}.
The phase shifts found by these two methods -- i.e., solving eq. (\ref{eq1a}) and the Euclidean space equation -- coincide with each other within 3-4 digits.
Furthermore, the imaginary part of the phase shifts vanishes with high accuracy below threshold. The unitarity condition is not automatically fulfilled in our approach, but appears as a consequence of handling the correct solution. It thus provides a stringent test of the numerical method. Our results reproduce the phase shifts given in  \cite{tjon} within the accuracy allowed by extracting numerical values from published figures.

\begin{figure}[btph]
\centering
\includegraphics[width=6.2cm]{fig1a.eps}
\hspace{0.3cm}
\includegraphics[width=7.cm]{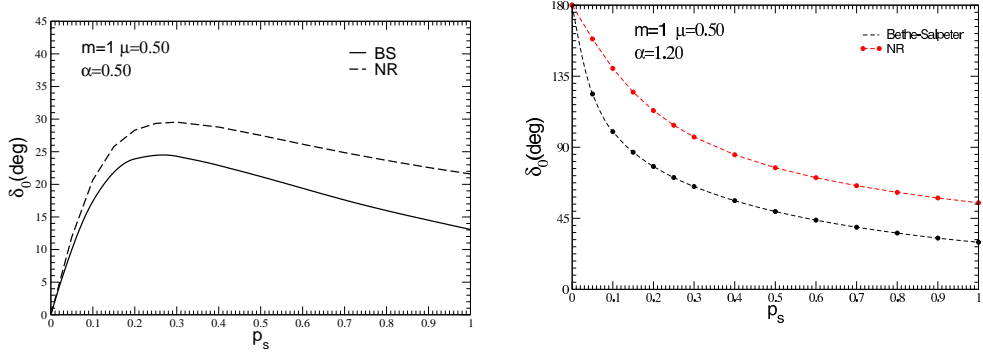}
\caption{Left panel: phase shift calculated  via BS equation (solid curve)  compared to  non-relativistic results  (dashed curve) for $\mu=0.5$ and $\alpha=0.5$.
Right panel: the same as for the left panel but for $\alpha=1.2$}
\label{fig1}
\end{figure}
Figure \ref{fig1} (left panel) shows the phase shifts calculated via BS equation (solid curve) and via the Schr\"odinger one with the Yukawa potential (dashed curve) for the constituent mass $m=1$,
exchange mass $\mu=0.5$ and coupling constant $\alpha=0.5$.
Right panel shows the same results for $\alpha=1.2$. For this value of $\alpha$, there exists a bound state. Therefore, according to the Levinson theorem,  the phase shift starts at 180 degrees.
One can see that the difference between relativistic and non-relativistic results is considerably large, specially  for small incident momentum. This difference increases with  $\alpha$.

\begin{figure}[btph]
\centering
\includegraphics[width=7.2cm]{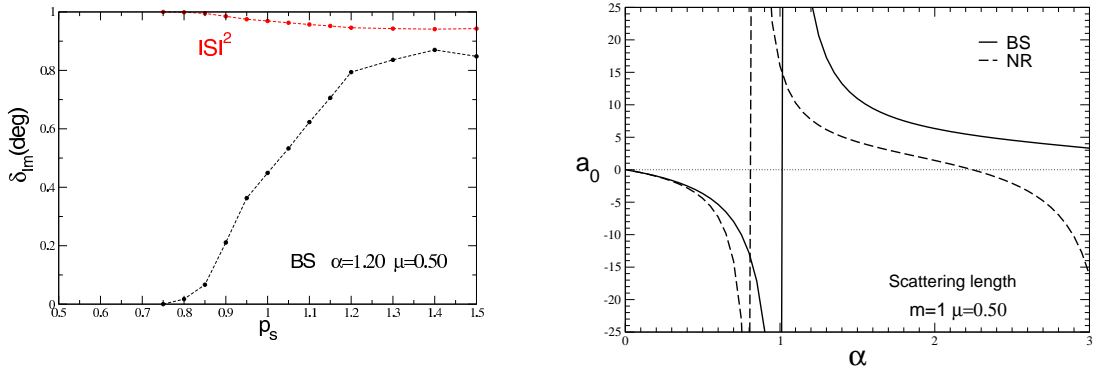}
\hspace{0.3cm}
\includegraphics[width=6.9cm]{fig2b.eps}
\caption{Left panel: imaginary part of the phase shift (solid)  and squared modulus of the S-matrix (dashed).
Right panel: scattering length $a_0$ vs. $\alpha$ for $\mu=0.5$ calculated via BS (solid) and   Schr\"odinger equations (dashed).}
\label{fig2}
\end{figure}

Fig. \ref{fig2} (left panel) shows the imaginary part of the phase shift
(for $\alpha=1.2$) which automatically appears when the incident momentum exceeds the threshold value for creation of the exchange meson. For $m=1$, $\mu=0.5$ this value $p^{thresh}=0.75$. Simultaneously, the modulas of two-body S-matrix differs from 1. For $p_s=1.118$ the second threshold, for creation of two mesons, is open. It also contributes in this curve.

Fig. \ref{fig2} (right panel)  shows the scattering length $a_0$ as a function of  the coupling constant $\alpha$ obtained with BS and non-relativsitc Schrodinger equations.
In the vicinity of  $\alpha\approx 0.8$ (for Schr\"odinger) and $\alpha\approx 1$ (for BS) the coupling constant crosses the critical value corresponding to the appearance of a bound state.
At this point the scattering length becomes infinite and then changes the sign.

\begin{figure}[btph]
\centering
\includegraphics[width=7.2cm]{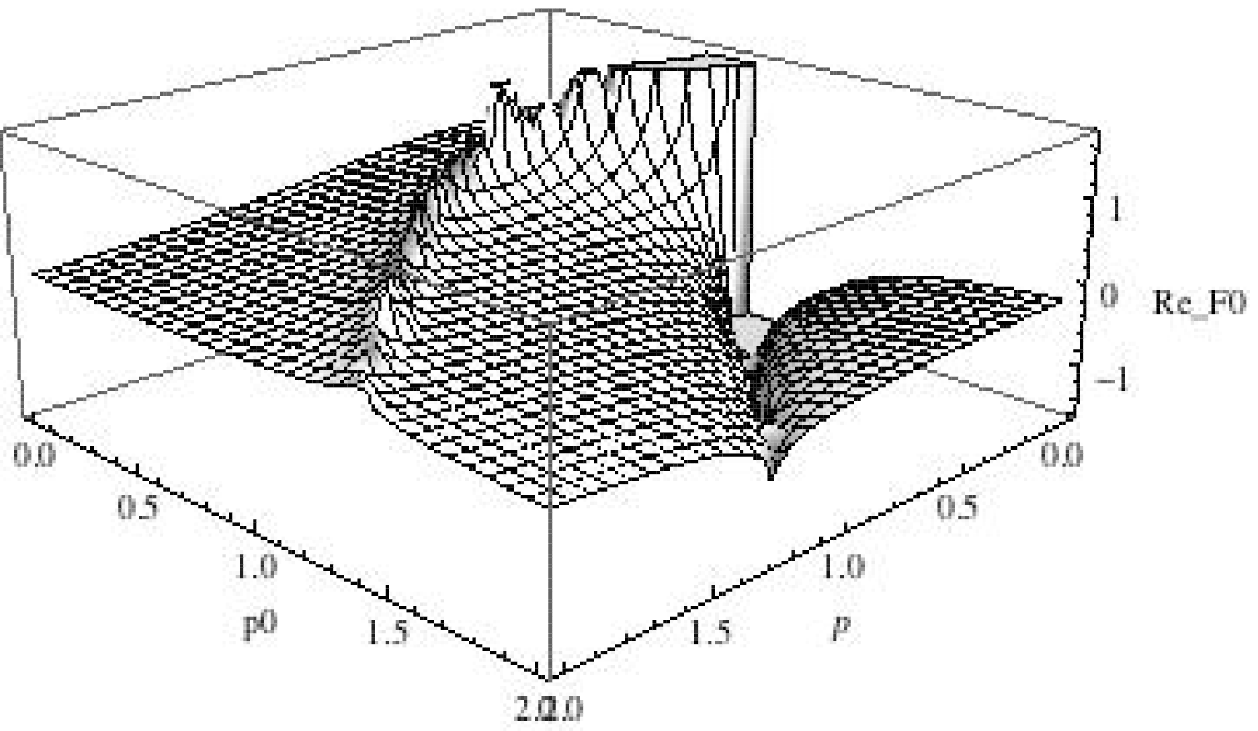}
\hspace{0.3cm}
\includegraphics[width=7.2cm]{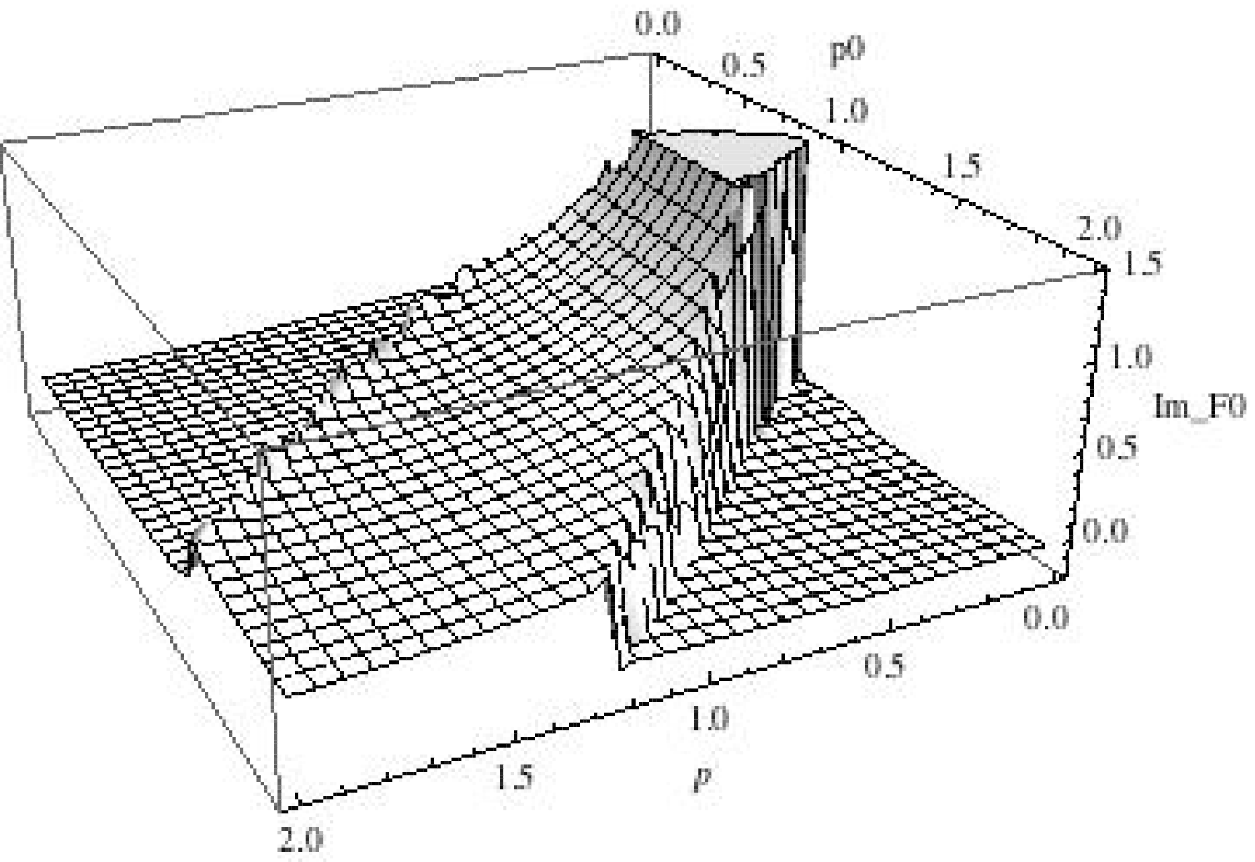}
\caption{Real (left) and imaginary (right) parts of the off-shell amplitude $F(p_0,p;p_s)$ for $p_s=0.5$, $\mu=0.5$.}
\label{fig3}       
\end{figure}

We have displayed in Fig. \ref{fig3} the real (left panel) and imaginary (right panel) parts of the off-shell scattering amplitude $F_0(p_0,p;p_s)$
vs. $p_0$ and $p$ calculated for $p_s=\mu=0.5$. Its real part shows a non trivial structure with  a ridge and a gap resulting from the singularities of the inhomogeneous term. Its on-shell value \mbox{$F_0^{on}=F_0(p_0=0,p=p_s;p_s)$}, determining the phase shift calculated previously, corresponds to a single point $p_0=0,p=p_s$ on these surfaces. Our calculation, shown in Fig. \ref{fig3}, provides  the full amplitude $F_0(p_0,p;p_s)$ in  a two-dimensional domain. It cannot be found from the Euclidean equation. Computing this quantity is the main result of this work.

\section{Conclusion}
We solved the BS equation for the scattering states in Minkowski space for the ladder kernel. The off-mass-shell amplitude is found for the first time. Coming on mass shell, we obtain the phase shifts which coincide with ones calculated by other methods. They considerably differ, even at low energy, from the non-relativisttic phase shifts calculated by the Schr\"odinger equation. Above the meson creation threshold the inelasticity appears which is also calculated. The off-mass-shell amplitude can be used to calculate the transition form factor and as an input in the three-body BS-Faddeev equations.



\end{document}